# Towards the characterization of individual users through Web analytics

Bruno Gonçalves[1] and José J. Ramasco[2]

[1] School of Informatics and Center of Biocomplexity, Indiana University,
Bloomington, IN 47408, USA
[2] Complex Systems and Networks Lagrange Laboratory (CNLL), ISI Foundation,
Turin, Italy
`bgoncalv@indiana.edu`,
WWW home page: `http://www.bgoncalves.com`

**Abstract.** We perform an analysis of the way individual users navigate in the Web. We focus primarily in the temporal patterns of they return to a given page. The return probability as a function of time as well as the distribution of time intervals between consecutive visits are measured and found to be independent of the level of activity of single users. The results indicate a rich variety of individual behaviors and seem to preclude the possibility of defining a characteristic frequency for each user in his/her visits to a single site.

## 1 Introduction

The introduction of *PageRank* by L. Page and S. Brin in 1999 and subsequent development of the Google search engine, marked a turning point in the history of the Internet [1]. For the first time a model of the way people interact with and navigate through Web pages was successfully applied to rank the content of the WWW. *PageRank* models web surfers as random walkers that blindly click through links without any memory of where they have been or where they intend to go. Despite the overwhelming success that resulted in one of the worlds fastest growing companies, it is clear that real users don't behave in the way prescribed by this simple set of rules. Several other models have been proposed over the years [2, 3, 4, 5, 6, 7, 8, 9, 10, 11, 12, 13, 14, 15, 16] but there has been a lack of data with which to test the merits or faults of each one. We analyze Web access log information from Emory University's server to provide a more accurate characterization of real world online behavior and how it evolves over time. Our final propose is twofold: on one hand, characterizing the navigation of users entering the domain and, on the other hand, studying the information that the logs contain about individuals' activity patterns. In the present work, we will focus mostly on the second question. Recently, there has been a surge of interest in the exploration of human dynamics through the daily interaction with a number of electronic devices [5, 17, 18, 19, 20, 21, 22, 23, 24, 25]. From mobile phones to social network sites; all provide abundant data on whom contacts whom and when it occurs. In a similar spirit but with a different perspective, we intend to study how users interact with Web pages. And more specifically, which are, if any, the rules that govern their return in time to the same site.

Our database is composed by the detailed logs generated by the software responsible for serving the content of Emory domain and respective sub domains during the period between Apr. 1, 2005 and Jan. 17, 2006. Due to privacy concerns, we have had access only to an anonymized version of these data. The sanitized version allows us to identify each user by an unique ID number, which is not traceable to the original IP. This ID number permits an accurate understanding of the behavior that each user displays in the online world, and the way accesses occur and evolve over time within the resolution of one second. One can think of this data set as a bipartite graph evolving in time. There are two types of "nodes": on one side, users, who connect to the other type representing URLs. Each link corresponds to a different request, that is timestamped with the date and time at which it occurred. During our observation period, the web site received over 3 million visitors to about 2.5 million pages for a grand total of over 53 million requests.

## 2 Number of visits

A Web domain with the size and variety of Emory's one inevitably attracts a fair amount of traffic. An important part of this attention, 32% of the total requests, are single time events. Many users as identified by their unique ID's never come back to the same Web page (URL) during the full period of our domain observation. This does not imply that they cannot return to other URLs. There can be different reasons for this behavior, among which the erratic Web exploration through search engines are likely to play a role. If one focus on the returning visitors alone, the statistics of the number of requests of a single user to an unique URL is quite broad. The histogram of the number of recurrent visits, $P(\omega)$, is displayed in Figure 1. The curve of $P(\omega)$ cannot be identified as a simple power-law (see the discussion in Ref. [17]) but is definitively long-tailed. Such broadness exposes a large variety in the users' behaviors. Variety that is not only constrained to the number of returning visits but also to their time location. In Figure 1b, the activity of a single user is shown. This user has been randomly selected among those visiting an administrative site that requires manual insertion of data, which ensures us that he or she is unequivocally human. The number of visits per unit of time shows drastic bursts followed by long periods of almost no activity. A similar bursty dynamics in time has been documented in mail and email answers and motivated the proposal of a model based on priority queuing [26, 27, 28, 29, 30]. We studied the application of this model to Web navigation in a previous work [17]. Next we will show extra details on the applicability of the model as well as new results regarding single user Web navigation.

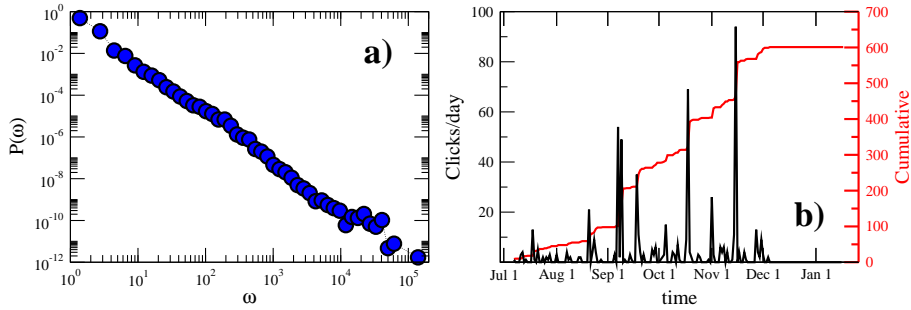

**Fig. 1.** In a), the histogram of the number of recurrent visits of a user to a single site. In b), activity profile of a single human user.

## 3 Time interval between consecutive visits

One most important question regarding single user behavior in our database is whether it is possible to distinguish automatic processes from genuine humans. The requests to a Web page can in fact be originated by a wide range of elements. Some that we were able to identify comprehend benign updating programs for software, malware, automatic hacker attacks and, finally, real human users. Most automatic processes are characterized by a very regular dynamics presenting a clear frequency, which makes them easy to detect and do not affect the return-time statistics. Intentional attacks however do not follow such predictable ways. Still, we are able to identify them by two characteristics: the huge amount of traffic they generate and its concentration in short periods of time. Plotting the average time between consecutive visits by a user to a page as function of the number of times he or she returns to that particular page as in Fig 2, a change of trend delates the presence of this type of malicious automatic processes. In the following, we will use these two identification techniques to filter out the contribution of automatic processes from the statistical study.

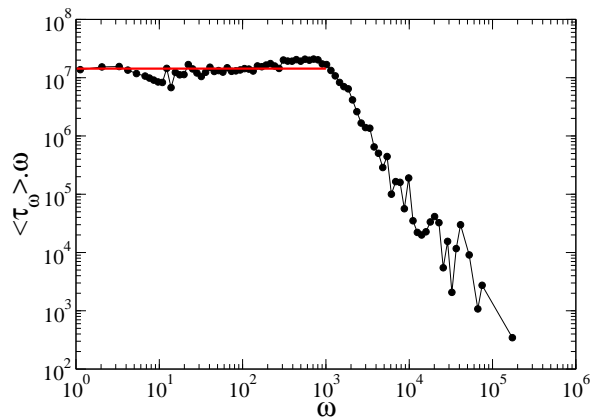

**Fig. 2.** Average inter-visit time of a single user to a single site $\langle \tau_\omega \rangle$ multiplied by $\omega$ as a function of the number of visits. A flat line (as the red curve) is expected for "standard" users, the change of trend after $w \sim 10^3$ indicates very different malicious activity.

In Ref. [17] we measured the time interval between two consecutive requests to a given web page by the same user. When measured over all the IP-URL pairs, it is clear that these intervals obey a power-law distribution with exponent $-1$ over about 5 orders of magnitude in time. This value of the exponent has important practical consequences. It implies that there is no statistically well defined average and that arbitrarily large fluctuations are possible, thus precluding the definition of a "normal" users. At first, this result strikes us as strange. After all, human beings are creatures of habit and many other studies in the social sciences have been able to define what "normal" behavior is in several situations. What makes web browsing different? Perhaps this peculiar exponent is simply an artifact of taking the average over several distinct classes of individuals. In Fig. 3 we plot this same distribution $P(\tau)$ but restricted to only a subset of the users. The first three curves are averaged only over users with less than 10, 100 and 1000 requests, respectively, over the course of the data sampling period. In all the cases the curve is identical to within small fluctuations and consistent with the previous $\tau^{-1}$ result. This implies that the users cannot be simply classified by their activity level and that, as we also showed in [31, 32], highly chaotic temporal patterns in the return times are an intrinsic characteristic of individual human users.

## 4 Return probability

As mentioned above, about 68% of requests made by all users are a consequence of a user returning to a page he or she has already visualized in the past. Intuitively, we would expect that the return probability $R(\tau)$ should vary over time. A hypothetical user will be less likely to revisit a page that has not seen for 2 years, than one that has visited it last week. If for no other reason than a simple loss of interests in the content, or even for having forgotten about a given resource even if it interested him. The way that this probability changes over time is, however, far from clear. To probe it, we have plot in Fig. 4 $R(\tau)$ as a function of the time $\tau$ after the user's first detected visit to a page. We can identify three different regimes corresponding to the three time scales: minutes, hours and days/months. In the first 10 minutes, users pressing the refresh or back button to reload the page are responsible to about 27% of all the returns. The return probability then quickly decreases until about 16 hours after the initial visit, accounting for roughly 33% of the total visits. This $16h$ period is consistent with students or staff returning to a page to finish some task started earlier in the day. The remaining 66% of returns can then be accounted for by people returning to the page to check for updated information. This regime is clearly different from the previous one, following a very slowly decreasing function that points to a much slower waning of interest at larger time scales.

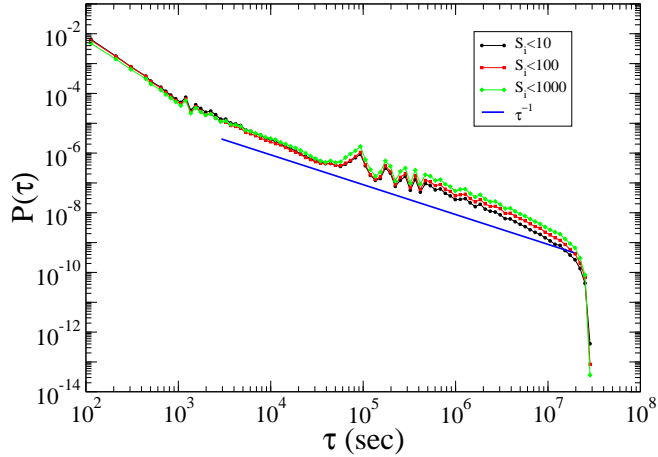

**Fig. 3.** Distribution of times between consecutive visits to a Web page by the same user thresholding the users by their level of activity.

As before for the users, we have extended our analysis to different groups of Web sites. In this case the discrimination is done by the popularity of the page, how many visits it accumulates during the whole period under study. As can be seen in Figure. 4b, the increase of the threshold in the number of visits does not perturb overall the functional shape of $R(\tau)$. It contributes however to a more nitid signal in the periods of time corresponding to entire days and a slightly longer decay of the tail. That is, the form of $R(\tau)$ as an estimator of how people loose interest on a site seems to be quite robust and relatively universal. More popular sites, as the home page of the university, show still a more marked influence from the returns produced within the usual circadian rhythms of human activity such as days and weeks.

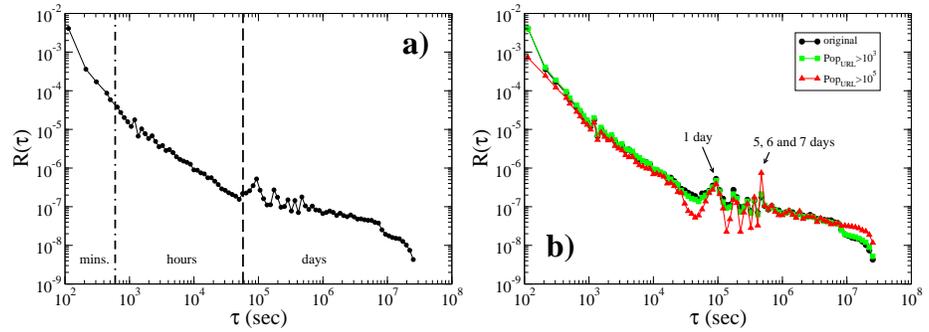

**Fig. 4.** Probability that a user will return to a page after time $\tau$. The curves on the right have the sites filtered by the number of visits they accumulated in our database, a measure of their popularity level.

## 5 Discussion

We have presented a detailed analysis of the user navigation patterns in the domain of a medium size American university. Our aim is to characterize better the propensity of users to revisit the same Web page and the distribution of times between such events. In order to do so, we estimated two functions: the probability of finding an inter-visit time of length $\tau$, $P(\tau)$, and the probability for a user of returning to the same site after a time $\tau$ of the first visit, $R(\tau)$. Both functions follow a very slow decrease with their respective time variables. $P(\tau)$ can be approximated by a power-law with exponent close to $-1$, while $R(\tau)$ does not display such a simple functional form being possible to recognize in it the different temporal scales. Our main finding is the robustness of both functions when the users or the Web sites considered for the analysis are filtered out accordingly to their activity or popularity. This fact points out to a possible universal behavior incarnated by their functional forms, and that would be an intrinsic characteristic of the relation between users and sites. More work, that is in progress, is needed to clarify how much this behavior extends to other different empirical systems out of work/academic environments.

## 6 Acknowledgements

This work was supported in part by grant NIH-1R21DA024259 from the National Institutes of Health. The authorss would like to thank A. Vespignani, S. Boettcher and the IT staff at Emory University for providing us with the log files used.

## References

1. Page, L., Brin, S., Motwani, R., Winograd, T.: The pagerank citation ranking: Bringing order to the web. Stanford Digital Library working paper SIDL-WP-1999-0120 (1999)
2. Gyongyi, Z., Garcia-Molina, H., Pedersen, J.: Combating web spam with trustrank. In *Proceedings of the Thirtieth international conference on Very large data bases - Volume 30*, (2004).
3. Radlinski, F., Joachims, T.: Active exploration for learning rankings from click-through data. In *ACM SIGKDD International Conference On Knowledge Discovery and Data Mining (KDD)*, (2007).

4. Liu, Y., Gao, B., Liu, T.-Y., Zhang, Y., Ma, Z., He, S., Li, H: Browserank: letting web users vote for page importance. In *Proceedings of the 31st annual international ACM SIGIR conference on Research and development in information retrieval*, (2008).

5. Meiss, M., Menczer, F., Fortunato, S., Flammini, A., Vespignani, A.: Ranking web sites with real user traffic. In *Proc. WSDM 2008*, (2008).

6. Watts, D.J.: A twenty-first century science. Nature 445, 489 (2007).

7. Pastor-Satorras, R., Vespignani, A.: Evolution and Structure of the Internet : A Statistical Physics Approach. Cambridge University Press, Cambridge, (2004).

8. Dorogovtsev, S., Mendes, J.F.F.: Evolution of Networks: From Biological nets to the Internet and WWW, Oxford University Press, Oxford (2003).

9. Watts, D.J., Strogatz, S.H.: Collective dynamics of "small-world" networks. Nature 393, 409–410 (1998).

10. Huberman, B.A., Pirolli, P.L., Pitkow, J.E., Lukose, R.M.: Strong regularities in World Wide Web surfing. Science 280, 95–97 (1998).

11. Barabási, A.-L., Albert, R.: Emergence of scaling in random networks. Science 286, 509–512 (1999).

12. Menczer, F.: Growing and navigating the small world Web by local content. Proc. Nat. Acad. Sci. 99, 14014–14019 (2004).

13. Dorogovtsev, S.N., Mendes, J.F.F.: Scaling properties of scale-free evolving networks:Continous approach. Phys. Rev. E 63, 056125 (2001).

14. Cattuto, C., Loreto, V., Servedio, V.D.P.: A Yule-Simon process with memory. Europhys. Lett. 76, 208–214 (2006).

15. Simkin, M.V., Roychowdhury, V.P.: A theory of Web traffic. EuroPhys. Lett. 82, 28006 (2007).

16. Wu, F., B.A. Huberman, B.A.: Novelty and collective attention. Proc. Nat. Acad. Sci. 104, 17599–17601 (2007).

17. Goncalves B., Ramasco, J.J.: Human dynamics revealed through web analytics. Phys. Rev. E 78, 026123 (2008).

18. Onnela, J.-P. *et al.*: Structure and tie strengths in mobile communication networks. Proc. Nat. Acad. Sci. 104, 7332–7336 (2007).

19. Golder, S., Wilkinson, D., Huberman, B.A.: Rhythms of social interaction: messaging within a massive online network. *3rd International Conference on Communities and Technologies* (CT2007). East Lansing, MI. June 28-30, 2007

20. Candia, J. *et al.*: Uncovering individual and collective human dynamics from mobile records. J. Phys. A 41, 224015 (2008).

21. Vázquez, A.: Impact of memory in human dynamics. Physica A 373, 747–752 (2007).

22. Brockmann, D., Hufnagel, L., Geisel, T.: The scaling laws of human travel. Nature 439, 462–465 (2006)

23. Zhou, T., Han, X.-P., Wang, B.-H.: Towards the understanding of human dynamics. `arXiv: 0801.1389` (2008).

24. Zhou, T., Kiet, H.-A.T., Kim, B.J., Wang, B.-H., Holme, P.: Role of activity in human dynamics. EuroPhys. Lett. 82, 28002 (2008).

25. González, M.C., Hidalgo, C.A., Barabási, A.-L.: Understanding individual human mobility patterns. Nature 453, 779–782 (2008).

26. Vázquez, A., *et al*: Modeling bursts and heavy tails in human dynamics, Phys. Rev. E 73, 036127 (2006).

27. Barabási, A.-L.: The origin of bursts and heavy tails in human dynamics. Nature 435, 207–211 (2005).

28. Oliveira, J.G., Barabási, A.-L.: Darwin and Einstein correspondence patterns. Nature 437, 1251 (2005).
29. Vázquez, A.: Exact results for the Barabási model of human dynamics. Phys. Rev. Lett. 95, 248701 (2005).
30. Oliveira, J.G., Vázquez, A.: Impact of interactions in human dynamics. `e-print ArXiv 0710.4916`, to appear in Physica A (2008).
31. Meiss, M., Duncan, J., Gonçalves, B., Ramasco, J.J., Menczer, F.: What's in a session: tracking individual behavior in the Web. Submitted to WWW 2009.
32. Gonçalves, B, Meiss, M., Ramasco, J.J., Flammini, A., Menczer, F.: Remembering what we like: Toward an agent-based model of Web traffic. Submitted to WSDM 2009